\documentclass[acmlarge]{acmart}

\usepackage{url}
\usepackage{booktabs} 
\usepackage{algorithmic}
\floatname{algorithm}{Procedure}
\usepackage{colortbl}
\usepackage[ruled]{algorithm2e} 

\usepackage{subcaption}
\usepackage{multirow}
\usepackage{tabularx}
\usepackage{bm}
\usepackage{graphicx}
\usepackage{rotating}
\usepackage{mathrsfs}

\SetAlFnt{\small}
\SetAlCapFnt{\small}
\SetAlCapNameFnt{\small}
\SetAlCapHSkip{0pt}
\IncMargin{-\parindent}

\acmJournal{JOCCH}
\acmVolume{9}
\acmNumber{4}
\acmArticle{39}
\acmYear{2017}
\acmMonth{3}
\acmArticleSeq{11}

\setcopyright{usgovmixed}

\acmDOI{0000001.0000001}

\begin{document}
\title{MindID: Person Identification from Brain Waves through Attention-based
Recurrent Neural Network}
 \author{Xiang Zhang}
 \affiliation{%
   \institution{University of New South Wales}
   \streetaddress{CSE, UNSW}
   \city{Sydney}
   \state{NSW}
   \postcode{2052}
   \country{AU}}

 \author{Lina Yao}
 \affiliation{%
   \institution{University of New South Wales}
   \streetaddress{CSE, UNSW}
   \city{Sydney}
   \state{NSW}
   \postcode{2052}
   \country{AU}}

   \author{Salil S. Kanhere}
 \affiliation{%
   \institution{University of New South Wales}
   \streetaddress{CSE, UNSW}
   \city{Sydney}
   \state{NSW}
   \postcode{2052}
   \country{AU}}

  \author{Yunhao Liu}
 \affiliation{%
   \institution{Tsinghua University}
   \streetaddress{School of Software, Tsinghua University}
   \city{Beijing}
   \state{Beijing}
   \postcode{100084}
   \country{China}}
   \author{Tao Gu}
 \affiliation{%
   \institution{RMIT University}
   \streetaddress{}
   \city{Melbourne}
   \state{VIC}
   \postcode{3001}
   \country{AU}}

 \author{Kaixuan Chen}
 \affiliation{%
   \institution{University of New South Wales}
   \streetaddress{CSE, UNSW}
   \city{Sydney}
   \state{NSW}
  \postcode{2052}
  \country{AU}}

\begin{abstract}
Person identification technology recognizes individuals by exploiting their unique, measurable physiological and behavioral characteristics. However, the state-of-the-art person identification systems have been shown to be vulnerable, e.g., anti-surveillance prosthetic masks can thwart face recognition, contact lenses can trick iris recognition, vocoder can compromise voice identification and fingerprint films can deceive fingerprint sensors.
EEG (Electroencephalography)-based identification, which utilizes the user's brainwave signals for identification and offers a more resilient solution, draw a lot of attention recently. However, the accuracy still requires improvement and very little work is focusing on the robustness and adaptability of the identification system.
We propose MindID, an EEG-based biometric identification approach, achieves higher accuracy and better characteristics.
At first, the EEG data patterns are analyzed and the results show that the Delta pattern contains the most distinctive information for user identification. Then the decomposed Delta pattern is fed into an attention-based Encoder-Decoder RNNs (Recurrent Neural Networks) structure which assigns varies attention weights to different EEG channels based on the channel's importance. The discriminative representations learned from the attention-based RNN are used to recognize the user' identification through a boosting classifier. The proposed approach is evaluated over 3 datasets (two local and one public). One local dataset (EID-M) is used for performance assessment and the result illustrate that our model achieves the accuracy of \textbf{0.982} which outperforms the baselines and the state-of-the-art. Another local dataset (EID-S) and a public dataset (EEG-S) are utilized to demonstrate the robustness and adaptability, respectively. The results indicate that the proposed approach has the potential to be largely deployment in practice environment.
\end{abstract}

%
%




\maketitle

\section{Introduction} 
\label{sec:Introduction}
Over the past decade, biometric information have been widely used in identification and have gained more acceptance due to their reliability and adaptability. Existing biometric identification systems are mainly based on individuals' unique intrinsic physiological features (e.g., face \cite{givens2013biometric}, iris \cite{latman2013field}, retina \cite{sadikoglu2016biometric}, voice \cite{goldstein2016methods}, and fingerprint \cite{unar2014review}). However, the state-of-the-art person identification systems have been shown to be vulnerable, e.g., anti-surveillance prosthetic masks can thwart face recognition, contact lenses can trick iris recognition, vocoder can compromise voice identification and fingerprint films can deceive fingerprint sensors.

The EEG (Electroencephalography) signal-based system is an emerging approach in physiological biometrics. Such systems measure an individual’s brain response to a number of stimuli in the form of EEG signals, which record the electromagnetic, invisible, and untouchable electrical neural oscillations. These characteristics enable EEG-based identification {\em highly attack-resilient} and escape from the threat of being deceived which is often faced by other identification techniques. For example, people can easily trick a fingerprint-based identification system by using a fake fingerprint film\footnote{\url{http://www.instructables.com/id/How-To-Fool-a-Fingerprint-Security-System-As-Easy-/}} or a face-recognition-based identification system by simply wearing a 200 dollars' worth anti-surveillance mask\footnote{\url{http://www.urmesurveillance.com/urme-prosthetic/}}. EEG signals, compared with other biometrics, have the following significant inherent advantages \cite{chuang2013think,sohankar2015bias}:
\begin{itemize}
  \item \emph {Attack-Resilience.} EEG data is invisible and untouchable and is impossible to be cloned and duplicated. Therefore, an EEG-based identification system is strengthened to verify human ID and robust against faked identities.

  \item \textit{Universality.} One's EEG signals are typically associated with the subject all the time and hence security can be enforced anywhere and anytime.

  \item \textit{Uniqueness.} Each individual processes his/her EEG signals which are unique, independent and different from other's \cite{gui2014exploring}. This can potentially achieve high identification accuracy.

  \item \textit{Accessibility.} We have seen an increasing effort in recent years in the development of low-cost and easy-to-wear EEG headsets. For example, the behind-the-ear EEG collection equipment\cite{kidmose2013study} can be easily attached to the ear (similar to wireless earphone).
\end{itemize}
We put up a table showing the comparison of EEG with other biometric information on several key characteristics in Table~\ref{tab:comparison_1}. EEG signals stand out in a number of aspects, and hence attract many research work in EEG-based biometric identification. For instance, Chuang et al. \cite{chuang2013think} propose a single-channel EEG-based identification system and Sarineh Keshishzadeh et al. \cite{keshishzadeh2016improved} employ a statistical model for analyzing EEG signals.

Despite the efforts done recently, the research work in EEG-based identification is still in its early stage, and several key challenges exist. One of the most significant challenges is poor stability (the identification system may work well at one time but fail another time due to the EEG signals are easy to be interfered). This may due to the user's physiological and psychological states such as fatigue and angry \cite{wang2015real,trejo2015eeg}. Intuitively, the states shift brought by the fluctuation of user states can be divided into two categories: the dramatically shift (e.g., hysterical, drunk, or under threaten) and the slight shift (e.g., headache or exciting). On one hand, the EEG signals divergence bought by the former can help to enhance the robustness of the identification system. For example, the phenomena could enhance the security of the identification system that it fails to recognize the subject who is under threatening (e.g., Hijacked by kidnapper). On the other hand, however, the latter will reduce the signal quality but more commonly occurred in the real world. For instance, the identification system could identify the user when s/he is happy but fail when upset. Thus the slight shift should be overcome for its negative effect. To eliminate the interference of the slight shift brought by the daily physical and mental states, we attempt to learn the robust and reliable representation via EEG pattern decomposition. Pattern decomposition is to decompose the full-frequency EEG signals into a specific pattern (Delta, Theta, Alpha, Beta, and Gamma).
 Pattern decomposition of EEG signals has been employed on EEG signal classification (e.g., movement task classification \cite{muller1999designing}) for a long time. However, few attentions are paid to the Delta pattern. In this paper, we discover that the Delta pattern is the most discriminative and efficient pattern through our analysis in Section~\ref{sec:eeg_bands_analysis}.

Another challenge is performance issue such as accuracy, robustness, and adaptability. The most recent identification systems can achieve a range from 80\% to 95\% \cite{marcel2007person,bashar2016human,keshishzadeh2016improved,kumari2015brainwave,thomas2016utilizing}, which is not enough for practical deployment in many confidential scenarios. Also, the identification algorithms rely much on the EEG collecting environment. The shifting of application environment (e.g., the number of channels, the sampling rate, and the training data size) may lead to the decrease of accuracy\footnote{This statement can be demonstrated in Section~\ref{sub:eeg_decomposition_effects}.}. This refers to that the existing EEG-based identification model may work well under one kind of application environment (e.g., 64 channels and 160 Hz), but could not handle another application environment (e.g., 14 channels and 128 Hz). So far, we have not seen a universal EEG-based identification algorithm which can performance good in a variety of real environments. To address this challenge, we introduce the attention-based RNNs (Recurrent Neural Networks) \cite{bahdanau2016end} which can automatically detect the most useful information from input data no matter what the environment is. More importantly, the attention mechanism\footnote{Simply, attention mechanism refers to select the most pertinent piece of information rather than using all available information. Attention Mechanisms in Neural Networks are based on the visual attention mechanism found in humans, and has been applied in computer version, NLP areas.} would automatically re-allocate the weights to extract most discriminative features according to the change of environmental factors. The efficiency of attention-based RNN framework has been demonstrated by the studies in areas such as speech recognition \cite{bahdanau2016end}, NLP (Natural Language Processing) \cite{ba2014multiple}, and computer version \cite{luong2015effective}.

To address the aforementioned problems, we propose MindID, a Delta pattern EEG-based person identification algorithm through an attention-based recurrent neural network.
Our main contributions of this paper are highlighted as follows:

\begin{itemize}
  \item We present an EEG-based identification approach, MindID, which adopts a novel attention-based Encoder-Decoder RNN framework for learning discriminative features among the user's brainwaves and utilizes the learned features to identify user ID through a boosting classifier. The attention mechanism enables our approach to automatically search the most discriminative features for identification, consequently, to operate robust and adaptive over different datasets and collecting environment.

  \item We analyze the EEG pattern decomposition and propose that the Delta pattern is the most steady and distinguishable pattern for user identification. Moreover, we design and conduct a set of experiments to verify the proposed hypothesis.

  \item We design and conduct an EEG experiment along with collecting two real-world local dataset (EID-M and EID-S) which are separately collected under single-trial and multi-trial\footnote{Single trial refers to that the dataset is collected in one session (the period from one subject putting the EEG headset on until all the experiment are finished then putting off). Multi-trials represents the EEG data is collected from different trials, which considered the effect on EEG data quality caused by the headset position errors.}.

  \item We evaluate the proposed approach on 3 datasets (2 local and 1 public). The results illustrate that our model achieves an accuracy of \textbf{0.982} which outperforms the state-of-the-art and baselines. We demonstrate the robustness and adaptability by the comparison between 3 datasets.

\end{itemize}

Note that all the necessary reusable codes and datasets in this paper have been open-sourced for reproduction, please refer to this link \footnote{\url{https://drive.google.com/open?id=1t6tL434ZOESb06ZvA4Bw1p9chzxzbRbj}}.


\begin{table}[]
\centering
\caption{Comparison of various biometrics. EEG have considerable rttack-resilient which is the most significant character of identification systems. $\uparrow$ denotes the higher the better while $\downarrow$ denotes the lower the better.}
\label{tab:comparison_1}
\resizebox{\textwidth}{!}{\begin{tabular}{llllllll}
\rowcolor[HTML]{C0C0C0}
\hline
  \textbf{Biometrics} & \textbf{Attack-Resilient} $\uparrow$ & \textbf{Universality} $\uparrow$ & \textbf{Uniqueness} $\uparrow$ & \textbf{Stability} $\uparrow$ &
  \textbf{Accessibility} $\uparrow$ & \textbf{Performance} $\uparrow$ & \textbf{Computational cost} $\downarrow$ \\ \hline
  Face/Vedio & Medium & Medium & Low & Low & High & Low & High \\
  Fingerprint/Palmprint & Low & High & High & High & Medium & High & Medium \\
  Iris & Medium & High & High & High & Medium & High & High \\
  Retina & High & Medium & High & Medium & Low & High & High \\
  Signature & Low & High & Low & Low & High & Low & Medium \\
  Voice & Low & Medium & Low & Low & Medium & Low & Low \\
  face & Medium & High & Medium & Medium & Medium & Medium & High \\
  Gait & High & Medium & High & Medium & Medium & High & Low \\
   {\bf EEG } & \textbf{High} & High & High & Low & Medium & High & Low \\ \hline
\end{tabular}
}
\end{table}

The remainder of this paper is organized as follows. Section~\ref{sec:Related Work} introduces the literature related to this paper. Section~\ref{sec:eeg_bands_analysis} analyzes the characteristics of EEG patterns. Section~\ref{sec:The Proposed approach} details the methodology of the MindID identification system. Section~\ref{sec:experiment_and_results} evaluates the proposed approach on the local and public dataset and provides analysis of the experimental results. Section~\ref{sec:discussion} discussed the limitation of our work and the future research potentials. Finally, Section~\ref{sec:conclusion} summarizes this paper and gives the conclusion.


\section{Related Work} 
\label{sec:Related Work}
In this section, we separately present literature on three aspects: the EEG-based person identification models, the EEG pattern decomposition, and the attention-based RNN application.
\subsection{EEG-based person identification} 
\label{sec:eeg_based}

Since EEG can be gathered in a safe and non-intrusive way, researchers have paid great attention to exploring this kind of brain signals. For person identification, EEG is promising for being confidential and attack-resilient but on the other hand, complex and hard to be analyzed. Marcel and Mill{\'a}n \cite{marcel2007person} use
Gaussian Mixture Models and train client models with Maximum A Posteriori (MAP). Ashby et al. \cite{ashby2011low} extract five sets of features from EEG electrodes and inter-hemispheric data, combine them together, and process the final features with support vector machine (SVM). The study shows that EEG identification is also feasible with less-expensive devices. Altahat et al. \cite{altahat2015analysing} select Power Spectral Density (PSD) as the feature instead of the widely used autoregressive (AR) models to get higher accuracy. They also conduct channel selection to determine contributing channels among all 64 channels. Thomas and Vinod \cite{thomas2016utilizing} take advantage of individual alpha frequency (IAF) and delta band signals to compose specific feature vector. They also prefer PSD features but only perform the extraction merely on gamma band. Most of the identification algorithms are concentrating on a specific application environment. Few studies attempt to build a universal EEG-based identification model.

\subsection{EEG pattern decomposition} 
\label{sub:eeg_pattern_decomposition}
 Generally, the EEG data could be decomposed into several patterns (delta, theta, alpha, beta, and gamma) corresponding to various brain states \cite{signalli2016}. So far, the majority of user ID identification studies are exploiting the features of Alpha and Beta pattern\cite{sohankar2015bias, kumari2015brainwave}. In particular, most EEG based identification models are work on the situation that the subject keeps rest/relax (under Alpha pattern) or concentrating state (under Beta pattern) for the high data quality. The rest and relax states are represented by the Alpha wave, therefore, a number of studies decompose EEG raw signals into the Alpha pattern for future analysis. Sohankar et al. \cite{sohankar2015bias} extract Alpha pattern features for identification and authentication. Bashar et al. \cite{bashar2016human} use the filtered signals with frequency ranges from $0.5-59 Hz$ (including Delta, Theta, Alpha, Beta and part of Gamma patterns) and calculate the statistics for user ID classification. Kumari and Vaish \cite{kumari2015brainwave} employ wavelet analysis to decompose original EEG signals into 5 patterns and extract statistical measures of each pattern. Thomas and Vinod \cite{thomas2016utilizing} take Alpha peak frequency and peak power and Delta band power as recognition features and achieves the highest recognition rate as 0.9. To our best knowledge, this paper is the very first work which specialized focus on the decomposition and analysis of Delta pattern and studies the person identification based on it (the justification is given in Section~\ref{sec:eeg_bands_analysis}).

\subsection{Attention-based RNN Model} 
\label{sub:attention_based_rnn_model}
Attention-based RNN model \cite{luong2015effective} refers to introduce attention mechanism to the RNN framework. The attention mechanism enables RNN algorithm to allocate different weights to different parts of the input, and consequently, improve the exploration of the corresponding relationship between the input sequence and the output sequence. Generally, attention module is added to the original RNN framework as an external module, however, the attention module is trained instantaneously with the RNN structure \cite{wang2016survey}. Attention-based RNN model has achieves success in speech recognition \cite{bahdanau2016end}, NLP (Natural Language Processing) \cite{ba2014multiple}, and computer version \cite{luong2015effective}. Bahdanau et al. \cite{bahdanau2016end} attempt to build a Large Vocabulary Continuous Speech Recognition (LVCSR) Systems by attention-based RNN and demonstrate this approach, compared with traditional methods, requires fewer training stages, less auxiliary data, and less domain expertise. Luong et al. \cite{ba2014multiple} explore the architecture of attention-based neural machine translation and exam the effects of two attentional mechanism (attends to all source words and attends to a subset of words) on the WMT translation tasks between English and German in both directions. Ba et al. \cite{luong2015effective} present an attention-based RNN model for recognizing multiple objects in images, which is attempts to recognize multiple objects despite being given only class labels during training. The results show that the attention-based RNN is more accurate and uses less computation than the state-of-the-art.
However, few work is taken based on attention mechanism in EEG related area. To our best knowledge, we are the very first work employing attention-based RNN model on the EEG-based user identification topic.

\section{EEG Pattern Analysis} 
\label{sec:eeg_bands_analysis}

\begin{table}[]
\centering
\caption{EEG patterns and corresponding characters. Awareness Degree denotes the awareness the degree of being aware of an external world.
}
\label{tab:bands}
\resizebox{\textwidth}{!}{\begin{tabular}{llllll}
\rowcolor[HTML]{C0C0C0}
\hline
\textbf{Patterns} & \textbf{Frequency ($Hz$)} & \textbf{Amplitude} & \textbf{Brain State} & \textbf{Awareness Degree} & \textbf{Produced Location} \\ \hline
\textbf{Delta} & 0.5-4 & Higher & Deep sleep pattern & Lower & Frontally and posteriorly \\
\textbf{Theta} & 4-8 & High & Light sleep pattern & Low &  Entorhinal cortex, hippocampus  \\
\textbf{Alpha} & 8-12 & Medium & Closing the eyes, relax state & Medium & Posterior regions of head \\
\textbf{Beta} & 12-30 & Low & Active thinking, focus, high alert, anxious & High & Most evident frontally \\
\textbf{Gamma} & 30-100 & Lower & During cross-modal sensory processing & Higher & Somatosensory cortex\\ \hline
\end{tabular}
}
\end{table}

In this section, we first introduce the basic knowledge of EEG patterns and then analyze the relationship between EEG patterns and individual states. Moreover, we propose a hypothesis to capture the most distinctive features to distinguish the subject's identity.

In practical EEG data analysis, the assembled EEG signals can be divided into several different frequency patterns (delta, theta, alpha, beta, and gamma) based on the strong intra-band correlation with a distinct behavioral state \cite{bacsar1980eeg,steriade1991alertness,signalli2016}. Each decomposed EEG pattern contains signals associated with particular brain information. The EEG frequency patterns and the corresponding characters are listed in Table~\ref{tab:bands}. The awareness degree denotes the perception of individuals while facing outside stimuli.
Each EEG patterns represents a specific active situation of brain state and a qualitative assessment of awareness. More specifically,
\begin{itemize}
\item {\bf Delta pattern} ($0.5-4$ Hz) is associated with deep sleep while the subject has lower awareness.
\item {\bf Theta pattern} ($4-8 $ Hz) being presented during light sleep, is the realm of the low awareness state.
\item {\bf Alpha pattern} ($8-12 $ Hz) mainly occurs during eye closed and deeply relax state, lies at the medium awareness.
\item {\bf Beta pattern} ($12-30 $ Hz) is the dominant rhythm while the subject keeps eye-opening and claims high awareness. Most of the human daily activities (such as eating, walking, and talking) are under Beta pattern.
\item {\bf Gamma pattern} ($30-100$ Hz) representing the co-work of several brain areas to carry out a specific motor and cognitive function. This pattern is associated with higher awareness.
\end{itemize}

\begin{table}[]
\centering
\caption{The inter-subject correlation coefficients. Full denotes the un-decomposed full-frequency band data. The lower coefficients indicate that the subject's EEG data is easier to be distinguished. The data come from the EID-M dataset (detailed in Section~\ref{sub:setting}).}
\label{tab:cc}
\resizebox{\textwidth}{!}{\begin{tabular}{llllllllllll}
\hline
\rowcolor[HTML]{C0C0C0}
{\bf Subject} & \textbf{} & \textbf{Subject 1} & \textbf{Subject 2} & \textbf{Subject 3} & \textbf{Subject 4} & \textbf{Subject 5} & \textbf{Subject 6} & \textbf{Subject 7} & \textbf{Subject 8} & \textbf{STD} & \textbf{Average} \\ \hline
 & Delta & 0.137 & 0.428 & 0.246 & 0.179 & 0.221 & 0.119 & 0.187 & 0.239 & 0.089554 & \textbf{0.219} \\
 & Theta & 0.447 & 0.671 & 0.552 & 0.31 & 0.387 & 0.207 & 0.199 & 0.386 & 0.151929 & 0.395 \\
 & Alpha & 0.387 & 0.629 & 0.615 & 0.377 & 0.299 & 0.306 & 0.283 & 0.457 & 0.128653 & 0.419 \\
 & Beta & 0.249 & 0.487 & 0.329 & 0.308 & 0.281 & 0.307 & 0.238 & 0.441 & 0.083224 & 0.33 \\
 & Gamma & 0.528 & 0.692 & 0.538 & 0.362 & 0.521 & 0.667 & 0.428 & 0.537 & 0.102288 & 0.534 \\
\multirow{-6}{*}{\textbf{Patterns}} & full & 0.333 & 0.329 & 0.408 & 0.304 & 0.297 & 0.621 & 0.302 & 0.447 & 0.104231 & 0.38 \\ \hline
\end{tabular}
}
\end{table}

We claim that {\em the EEG patterns are internally related with the awareness degree}. As shown in Table~\ref{tab:bands}: with the increase of band frequency (from Delta pattern, Theta pattern, Alpha pattern, Beta pattern to Gamma pattern),  the awareness degree is increasing.  The above statement can be inferred by the following two factors. First, EEG pattern is relevant to brain neuron activity. In essence, EEG signals are measured by the voltage fluctuations which are resulted from the ionic current within the neuron activity of the brain \cite{moruzzi1949brain}.
Second, the awareness degree is associated with the brain neuron activity. Intuitively, the higher awareness the subject has, the more neurons are activated. In particular, more and more brain areas are activated while the subject's  awareness is higher and higher (the brain state changes from deep sleep, light sleep, to normal awake). At the same time, more neural cells are aroused and more function are attached. As a result, more complex and blend EEG signals are produced by the brain.

Additionally, we know that {\em the awareness of human is naturally connected with individuals' mental and physical states (organics and systems)} \cite{edwards2013pacing,noakes2011time}. For example, while the subject is under lower awareness situation (like deep sleep, Delta pattern), the most parts of physical functions of the body (such as sensing, thinking, even dreaming) are completely detached. Only the very essential life-support organs and systems (such as breathing, heart beating, and digesting) keep working.
While the subject is under medium awareness state (like eye relaxation, Alpha pattern), the subject has more activated functions such as imaging, visualizing and concentrating. Also, more brain functions like hearing, touching, and thinking are attached, which means that more physical brain areas (such as frontal lobe, temporal lobe, and parietal lobe) are activated.

According to the above two statements, it can be inferred that {\bf the EEG patterns are associated with individuals' mental and physical states (organics and systems)}.
Note, under the medium awareness situation, the life-support systems which worked under lower awareness situation are still working. Which means that while the subject has high-degree awareness, his or her EEG signals contain both high-degree and low-degree awareness at the same time. { The pattern (with low-degree awareness) is {\bf not replaced} by another pattern (with high-degree awareness) but {\bf included} by the latter}. Specifically, Delta pattern is not replaced but included in other patterns. In other word, Delta pattern exists in all the brain states\footnote{For example, the deep sleep EEG signals contain only Delta pattern wile the light sleep EEG signals contain both Delta and Theta patterns.} (e.g., deep sleep, light sleep, relax, and focus).

For identification techniques, the EEG signals feature should satisfy two demands: steady and distinguishable. The steady means that the system should be robustness enough to identify the user even when the user's mental or physical states have tiny fluctuation (such as tired). The distinguishable refers that the EEG signals should vary with the different subject. Based on the analyzed conclusion, we claim a hypothesis that {\bf Delta pattern contains the most steady and distinctive information for user identification}. This hypothesis will be demonstrated both qualitatively and quantitatively.

Here we attempt to {\em qualitatively demonstrate the hypothesis} based on the relationship between EEG patterns and human states. At first, Delta pattern naturally keeps steady under different situations since it is produced by and only related with the basic life-support systems. Comparatively, other EEG patterns like the Alpha pattern is unsteady and it could be easily influenced by subject states and environmental factors (such as fatigue, emotion, and noise). The higher consciousness, the easier to be influenced by the noise. In addition, the life-support systems are associated with the physiological characteristics of different subjects, which enables Delta pattern distinguishing.
Then we present {\em the quantitative demonstration}. To find the best pattern for user ID recognition, we analyze the inter-subject correlations of EEG decomposed pattern. The inter-subject correlations, measured by the correlation coefficient, denotes that the connection of the same pattern but from different subjects. For example, the inter-subject Alpha pattern correlations of subject 1 are calculated by the Pearson correlation coefficient between the Alpha signal (belong to subject 1) and another Alpha signal (belong to another subject). In practical, we select a set of samples and measure the average level. The correlation coefficient analysis results are shown in Table~\ref{tab:cc}. In which, we can observe that the Delta pattern has the lowest correlation coefficients compared with other patterns. This consequence indicates that Delta pattern is enabled to achieve the best performance for the user identification.
Furthermore, the comparative experiment between different EEG patterns will be reported in Section~\ref{sub:eeg_decomposition_effects}.


\section{Methodology} 
\label{sec:The Proposed approach}
In this section, we first give an overview of the proposed MindID system and then present the technical details for each component, namely, {\em Preprocessing}, {\em EEG pattern decomposition}, {\em Attention-based RNN}, and {\em Classification}.
\begin{figure}
  \includegraphics[width=\linewidth]{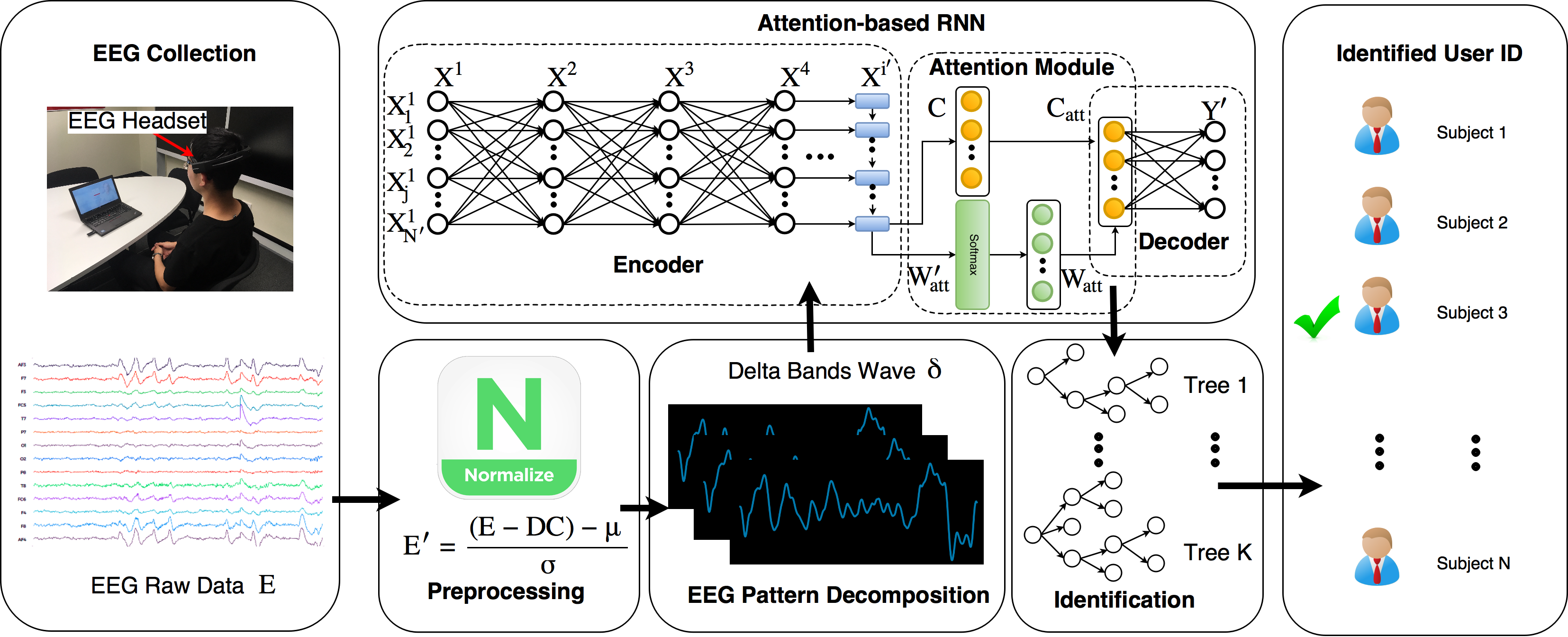}
  \caption{Flowchart of the proposed approach. In the beginning of identification, EEG raw data $E$ is collected from the user and then be transmitted to preprocessing stage. The preprocessed data $E'$ is decomposed to Delta pattern $\delta$ which is regarded as the input of the attention-based RNN. The encoder compresses the input sequence $X^1$ into an intermediate coder $C$ and produces the weights $W'_{att}$ at the same time. The attention-based module accepts both $C$ and $W'_{att}$ from the LSTM layer $X^{i'}$, processes $W'_{att}$ through a softmax layer, and calculates the attention-based code $C_{att}$. Assess the representation ability of $C_{att}$ via the decoder and utilize it to identify the user ID in the identification step.
  }
  \label{fig:workflow}
\end{figure}

\subsection{Overview} 
\label{sub:overview}
Figure~\ref{fig:workflow} outlines the specific steps of the proposed MindID system.
The brainwave is collected by the portable EEG acquisition equipment while the user {\it closed his/her eyes and keep relaxation}. Under relaxation mental and physical states, the EEG signals are supposed to be more stable and reliable.
Each EEG data is a numerical feature vector with N dimensions which corresponding to the N channels of the wearable EEG headset.
The EEG samples are first preprocessed to remove the Direct Current (DC) offset and normalize the signals (Section~\ref{sub:preprocessing}). Next, we employ EEG pattern decomposition to isolate the Delta waves from preprocessed data since they contain the most distinctive information which can be used to identify the subject (as outlined in Section~\ref{sec:eeg_bands_analysis}). The delta waves are fed to an attention-based Encoder-Decoder RNN model, which identifies the most distinctive channels and adjusts the weights accordingly.
The attention-based RNN model accepts Delta pattern signals and explores the deep correlations between Delta pattern. The learned deep representations are fed into a statistical boosting classifier (Section~\ref{sub:identification}) to recognize the user ID.


\subsection{Preprocessing} 
\label{sub:preprocessing}
The raw EEG samples are pre-processed to remove the DC offset and normalize the signals.

Eliminating DC offset is necessary because EEG collection headsets invariably introduce a constant noise component in the recorded signals. The specific headset used in our experiments (details in Section~\ref{sec:experiment_and_results}) introduces a DC offset of  4200 muV\footnote{\url{https://www.bci2000.org/mediawiki/index.php/Contributions:Emotiv}}. In the preprocessing stage, this constant DC offset is first subtracted from the raw signal E.


Normalization also plays a crucial role in a knowledge discovery process for handling different units and scales of features. For example, given one raw data dimension ranges from 0 to 1 while another dimension ranges from 0 to 100, the analysis results will be dominated by the latter. Generally, there are three widely used normalization methods: Min-Max Normalization, Unity Normalization, and Z-score Scaling Normalization \cite{zhang2017multi}.
Our experiments (not shown for brevity) indicated that Z-score scaling is the most suited for the EEG data. In summary, the preprocessed data $E'$ can be calculated by
$$E'=\frac{(E-DC)-\mu }{\sigma }$$
where $DC$ denotes the Direct Current which is 4200 muV, $\mu$ denotes the mean of $E-DC$ and $\sigma$ denotes the standard deviation.


\subsection{EEG Pattern Decomposition} 
\label{sub:eeg_bands_decomposition}
In Section 3, we used empirical EEG data to show that the part of the EEG signals that belong to the Delta frequency band ($0.5-4 Hz$) is particularly well-suited to identify user's ID accurately and steady. To isolate the signals in the Delta band, we use a Butterworth band-pass filter of order 3 with the frequency range of 0.5Hz to 4Hz.
The designed filter has following specifications: the order is three, the low cut is $0.5 Hz$, and the high cut is set as $4 Hz$.
All dimensions of the preprocessed $E'$ are fed into the band-pass filter in turn and finally get the decomposed Delta pattern $\delta$.


\begin{table}%
\caption{Notation}
\label{tab:notation}
\begin{minipage}{\columnwidth}
\begin{center}
\begin{tabular}{ll}
  \toprule
  \bf Parameters& \bf Explanation\\
  \hline
  $E$     & EEG raw data\\
  $E'$     & Preprocessed EEG data\\
  $\delta$     & Delta pattern of $E'$\\
  $X^i$     & Data in the $i$-th layer in attention-based RNN\\
  $I$   &The number of layers in attention-based RNN\\
  $N^i$   &The number of dimensions of $X^i$\\
  $Y$     & The one-hot label of user ID\\
  $Y'$     & The attention-based RNN predicts user ID\\
  $K$ & The number of user ID categories\\
  $\mathcal{T}(\cdot)$ & The linear function\\
  $C$     & The intermediate code\\
  $\mathcal{L}(\cdot)$     & The output calculation procedure of LSTM cell\\
  $\mathcal{L}'(\cdot)$     & The final hidden state calculation procedure of LSTM cell\\
  $f_i,f_f,f_o,f_m$          & The input, forget, output, and input modulation gate\\
  $W'_{att}$      &  The unnormalized attention weights\\
  $W_{att}$ & The normalized attention weights\\
  $C_{att}$ & The attention-based intermediate code\\
  $n_{iter}$ & The iteration threshold of attention-based RNN\\
  $X_D$ & The learned deep feature from attention-based RNN\\
  $x_d$ & A single sample in $X_D$\\
  $m$ &The $m$-th tree \\
  $M$ & The number of XGB trees\\
  $I_D$ & The final identified user ID of MindID approach\\
  \bottomrule
\end{tabular}
\end{center}
\end{minipage}
\end{table}%

\subsection{Attention-based RNN} 
\label{sec:deep_feature_learning}
After EEG pattern decomposition, the composed Delta pattern $\delta$ is fed into an attention-based Encoder-Decoder RNN structure \cite{wang2016survey} aims to learn more representable features for user identification. The general Encoder-Decoder RNN framework regards all the feature dimensions of input sequence has the same weights, no matter how important the dimension is for the output sequence. In our research, the different feature dimensions of the EEG data are corresponding to the different nodes of the EEG equipment. For example, the first dimension (first channel) collects the EEG data from the $AF3$\footnote{Both $AF3$ and $O2$ are EEG measurement positions in the International 10-20 Systems.} node which located at the frontal lobe of the scalp while the 7-th dimension is gathered from $O1$ node at the occipital lobe. To assign varies weights to different dimensions of the brainwave data, we introduce the attention mechanism to the Encoder-Decoder RNN model. The proposed attention-based Encoder-Decoder RNN is consists of three components (as shown in Figure~\ref{fig:workflow}): the encoder, the attention module, and the decoder. The encoder is designed to compress the input Delta $\delta$ wave into a single intermediate code $C$; the attention module helps the encoder to calculate a better intermediate code $C_{att}$ by generating a sequence of the weights $W_{att}$ of different dimensions; the decoder accepts the attention-based code $C_{att}$ and decode it to the user ID. Note, this user ID is predicted by the attention-based RNN instead of MindID, and the final identified ID of MindID approach will be introduced in Section~\ref{sub:identification}.

Suppose the data in $i$-th layer could be denoted by $X^i=(X^i_j;i\in[1,2,\cdots,I], j\in[1,2,\cdots, N^i])$
where $j$ denotes the $j$-th dimension of $X^i$. $I$ represents the number of neural network layers in the proposed attention based RNN model while $N^i$ denotes the number of dimensions in $X^i$. Take the first layer as an example, we have $X^1=\delta$ which indicates the input sequence is the Delta pattern. Let the output sequence be $Y=(Y_k; k\in[1,2,\cdots,K])$ where K denotes the number of user ID categories. In this paper, the user ID is represented by the one-hot label with length $K$.
For simplicity, let's define the operation $\mathcal{T}(\cdot)$ as:
$$\mathcal{T}(X^i)=X^iW+b$$
Further more, we have
$$\mathcal{T}(X^{i-1}_j,X^i_{j-1})=X^{i-1}_j*W'+X^i_{j-1}*W''+b'$$
where $W$, $b$, $W'$, $W''$, $b'$ denote the corresponding weights and biases parameters.

The the encoder component contains several non-recurrent fully-connected neural network layers and one recurrent Long Short-Term Memory (LSTM) layer. The non-recurrent layers are employed to construct and fit a non-linear function to purify the input Delta pattern, the necessity is demonstrated by the preliminary experiments\footnote{Some optimal designs like the neural network layers are validated by the preliminary experiments but the validation procedure will not be reported in this paper for space limitation}. The data flow in these non-recurrent layers could be calculated by
$$X^{i+1}=\mathcal{T}(X^i)$$
The LSTM layer is adopted to compress the output of non-recurrent layers to a length-fixed sequence which is regarded as the intermediate code $C$. Suppose LSTM is the $i'$-th layer, the code equals to the output of LSTM, which is $C=X^{i'}_j$. The $X^{i'}_j$ can be measured by
\begin{equation}
\label{equ:1}
X^{i'}_j=\mathcal{L}(c^{i'}_{j-1},X^{i-1}_j,X^{i'}_{j-1})
\end{equation}
 where $c^{i'}_{j-1}$ denotes the hidden state of the $(j-1)$-th LSTM cell. The operation $\mathcal{L}(\cdot)$ denotes the calculation process of the LSTM structure, which can be inferred from the following equations
 \[X^{i'}_{j}=f_o\odot tanh(c^{i'}_{j})\]
  \[c^{i'}_{j}=f_f\odot c^{i'}_{j-1}+f_i\odot f_m\]
 \[f_o=sigmoid(\mathcal{T}(X^{i'-1}_{j},X^{i'}_{j-1}))\]
\[f_f=sigmoid(\mathcal{T}(X^{i'-1}_{j},X^{i'}_{j-1}))\]
 \[f_i=sigmoid(\mathcal{T}(X^{i'-1}_{j},X^{i'}_{j-1}))\]
\[f_m=tanh(\mathcal{T}(X^{i'-1}_{j},X^{i'}_{j-1}))\]
where $f_o,f_f, f_i$ and $f_m$ represent the output gate, forget gate, input gate and input modulation gate\footnote{\url{http://colah.github.io/posts/2015-08-Understanding-LSTMs/}},
separately, and $\odot$ denotes the element-wise multiplication.

The attention module accepts the final hidden states as the unnormalized attention weights $W'_{att}$ which can be measured by the mapping operation $\mathcal{L}'(\cdot)$ (similar with Equation~\ref{equ:1})
$$W'_{att}=\mathcal{L}'(c^{i'}_{j-1},X^{i-1}_j,X^{i'}_{j-1})$$
and calculate the normalized attention weights $W_{att}$
$$W_{att}=softmax(W'_{att})$$
The softmax function is employed to normalize the attention weights into the range of $[0,1]$. Therefore, the weights can be explained as the probability that how the code $C$ is relevant to the output results.
Under the attention mechanism, the code $C$ is weighted to $C_{att}$
$$C_{att}=C\odot W_{att}$$
Note, $C$ and $W_{att}$ are trained instantaneously.
The decoder receives the attention-based code $C_{att}$ and decode it to predict the user ID $Y'$\footnote{Note, $Y'$ is not the identification results of MindID model. The final identified user ID is $I_D$ calculated in Section~\ref{sub:identification}} . Since $Y'$ is predicted at the output layer of the attention based RNN model ($Y'=X^I$), we have
$$Y'=\mathcal{T}(C_{att})$$
At last, we employ the cross-entropy function to calculate the prediction cost between the predicted ID $Y'$ and the ground truth $Y$. $\ell_2$-norm (with parameter $\lambda$) is selected to prevent overfitting. The cost is optimized by the AdamOptimizer algorithm \cite{kingma2014adam}. The iterations threshold of attention-based RNN is set as $n_{iter}$. The weighted code $C_{att}$ has a directly linear relationship with the output layer and the predict results. If the model is trained well and get low cost, we could regard the weighted code as a high-quality representation of the user ID. We set the learned deep feature $X_D$ equals to $C_{att}$, $X_D=C_{att}$, and use it to recognize the final user ID in the identification stage.


\subsection{Identification} 
\label{sub:identification}

In this section, we employ Extreme Gradient Boosting classifier (XGB) \cite{chen2016xgboost} to classify the learned deep feature $X_D$ for user ID identification. The XGB classifier fuses a set of classification and regression trees (CART) and tries to exploit as detailed as possible the information from the input data.
It builds multiple trees and each tree has its leaves and corresponding scores. Moreover, it proposes a regularized model formalization to prevent over-fitting and it is widely used for its accurate prediction power.

The learned deep feature $X_D$ is taken to train a list of the CART (set there are $M$ trees) and predict a set of user's IDs. Suppose $x_d\in X_D$ is a single sample of the deep feature. The finally identification result of the input $x_d$ is calculated as
$$y_m=f(x_d)$$
$$I_D=F(\sum_1^M y_m), m=1,2, \cdots,M$$
where $f$ denotes the classify function of a single tree, $y_m$ denotes the predicted ID of the $m$-th tree and $F$ denotes the mapping from single tree prediction space to the final prediction space. The $I_D$ is the final identified user ID based on EEG data.
The overall procedure of the proposed approach is summarized in Algorithm~\ref{alg:approach}. All the parameters mentioned in this section are listed in Table~\ref{tab:notation}.

\begin{algorithm}[!t]
\caption{The MindID User Identification Algorithm}
\label{alg:approach}
\begin{algorithmic}[1]
\renewcommand{\algorithmicrequire}{\textbf{Input:}}
\renewcommand{\algorithmic}{\textbf{Hyper-parameters:}}
 \renewcommand{\algorithmicensure}{\textbf{Output:}}
 \REQUIRE EEG raw data $E$
 \ENSURE  Identification results $I_D$
 \STATE  Initialization\;
 \STATE Preprocessing: $E'\leftarrow E$\;
 \STATE EEG pattern decomposition: $\delta \leftarrow E'$\;

  \IF{$iteration<n_{iter}$}
      \FOR{$i=1,2,\cdots,I$}
        \STATE $X^1=\delta$
        \STATE $C \leftarrow X^1, \mathcal{L}(c^{i'}_{j-1},X^{i-1}_j,X^{i'}_{j-1})$
        \STATE $W_{att} \leftarrow C, \mathcal{L}'(c^{i'}_{j-1},X^{i-1}_j,X^{i'}_{j-1})$
        \STATE $C_{att}=C\odot W_{att}$
        \STATE $X_D=C_{att}$
      \ENDFOR
  \ELSE
      \STATE Return $X_D$
  \ENDIF
  \FOR{$X_D$}
    \STATE $I_D\leftarrow X_D$
  \ENDFOR
\RETURN $I_D$
\end{algorithmic}
\end{algorithm}

\begin{figure}[h!]
    \centering
    \includegraphics[width=\textwidth]{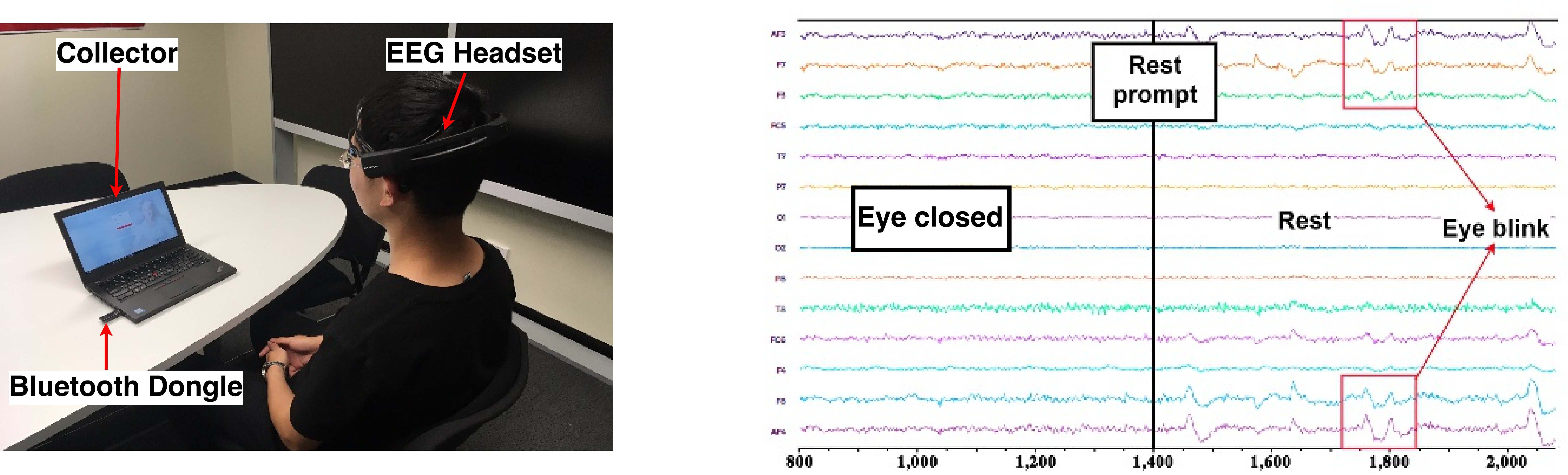}
    \caption{EEG collection and the collected raw data. The EEG raw data is gathered by the EEG headset and transmitted to the collector through bluetooth. The EEG data with the user keep relaxation and eye-closed is recorded.}
    \label{fig:EEG_collection}
\end{figure}

\section{Experiments and results} 
\label{sec:experiment_and_results}
We first outline the experimental settings in Section~\ref{sub:setting}. Next, we systematically investigate the following questions:
\begin{itemize}
\item How does MindID compare with state-of-the-art methods and other baselines (Section~\ref{sub:overall_comparison})?
\item How efficient is MindID (Section~\ref{sub:efficiency_evaluation})?
\item Is MindID robust under a multi-trial setting (Section~\ref{sub:robustness_evaluation})?
\item Does MindID exhibit consitent results when tested with different datasets (Section~\ref{sub:adaptability_evaluation})?
\item Is the Delta pattern exactly works better than other patterns (Section~\ref{sub:eeg_decomposition_effects})?

\end{itemize}

\subsection{Experimental settings} 
\label{sub:setting}

\subsubsection{Datasets} 
\label{sub:datasets}
The proposed MindID system is evaluated by three datasets: a multi-trial local dataset ({\it EID-M}), a single-trial dataset ({\it EID-S}), and a public dataset ({\it eegmmidb}). The details of datasets are introduced in Table~\ref{tab:datasets}.
All the datasets are measured the EEG raw data from the subject's scalp while the subject keeps {\it relax and eye-closed}.

\begin{table}[]
\centering
\caption{Datasets details. In Trial column, M denotes multi-trials and S demotes single-trial. {\bf EID-M} is used to compare with the state-of-the-arts and baselines; the comparison between {\bf EID-M} and {\bf EID-S} are used to verify the robustness; the comparison between {\bf EID-S} and {\bf EEG-S} are used to verify the adaptability.}
\label{tab:datasets}
\resizebox{\textwidth}{!}{\begin{tabular}{lcccccccc}
\rowcolor[HTML]{C0C0C0}
\hline
\textbf{Name} & \textbf{Source} & \textbf{Channels} & \textbf{Trial} & \textbf{Frequency} & \textbf{Subjects} & \textbf{Comparison} & \textbf{Robustness} & \textbf{Adaptability} \\ \hline
\textbf{EID-M} & Local & 14 & M & 128 $Hz$ & 8 &\checkmark  & \checkmark   & - \\
\textbf{EID-S} & Local & 14 & S & 128 $Hz$ & 8 &       -       & \checkmark &\checkmark  \\
\textbf{EEG-S} & Public & 64 & S & 160 $Hz$ & 8 & -     &     -      &\checkmark \\ \hline
\end{tabular}
}
\end{table}

\textbf{EID-M} EID-M denotes EEG based ID recognition with the training set comes from the multi-trial collection. Since multi-trial scenarios are mostly happed in the practical applications, EID-M dataset is taken to report a comparison with the state-of-the-art methods and baselines.
The EID-M dataset is gathered in the experiment which is carried on by 8 subjects (5 males and 3 females) aged from 24 to 28. During the experiment, the subject wearing the \textit{Emotiv Epoc+}\footnote{\url{https://www.emotiv.com/product/emotiv-epoc-14-channel-mobile-eeg/}} EEG collection headset, keeping relax and eye-closed (shown in Figure~\ref{fig:EEG_collection}). The Emotiv Epoc+ contains 14 channels and the sampling rate is set as 128 $Hz$. In the experiment, each subject takes three trials and each trial produce 7,000 EEG samples. Summarily, each subject has 21,000 samples and the whole EID-M dataset contains 168,000 samples.

\textbf{EID-S} EID-S is collected under the same situation with EID-M (5 males, 3 females, 14 channels, and 128 Hz). The main difference between them is the former dataset are belonged single trial. EID-S totally contains 56,000 samples belong to 8 subjects (7,000 samples belong to each subject).

\textbf{EEG-S} EEG-S is a subset of the widely used online public dataset {\it eegmmidb} (\textit{EEG motor movement/imagery database})\footnote{\url{https://www.physionet.org/pn4/eegmmidb/}}. It is collected with the BCI2000 (Brain Computer Interface) instrumentation system \footnote{\url{http://www.schalklab.org/research/bci2000}} \cite{schalk2004bci2000} (64 channels and 160 $Hz$ sampling rate). EEG-S contains 8 subjects with each subject owns 7,000 samples which are collected in single trial.

To assess the performance of the proposed MindID model, we employ several widely-used evaluation metrics such as accuracy, precision, recall, F1 score, ROC (Receiver Operating Characteristic) curve, support, and AUC (Area Under the Curve).

\begin{figure}[ht]
\centering
\begin{minipage}[b]{0.32\linewidth}
\centering
\includegraphics[width=\textwidth]{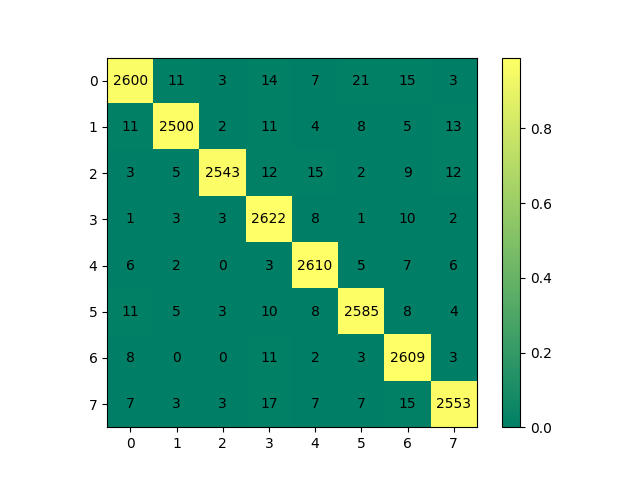}
\caption{Confusion matrix of EID-M}
\label{fig:con_m_3trials}
\end{minipage}
\begin{minipage}[b]{0.32\linewidth}
\centering
\includegraphics[width=\textwidth]{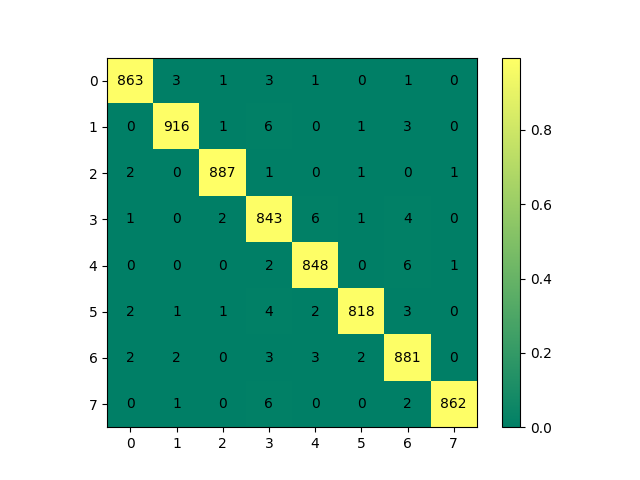}
\caption{Confusion matrix of EID-S}
\label{fig:con_m_1trial}
\end{minipage}
\begin{minipage}[b]{0.32\linewidth}
\centering
\includegraphics[width=\textwidth]{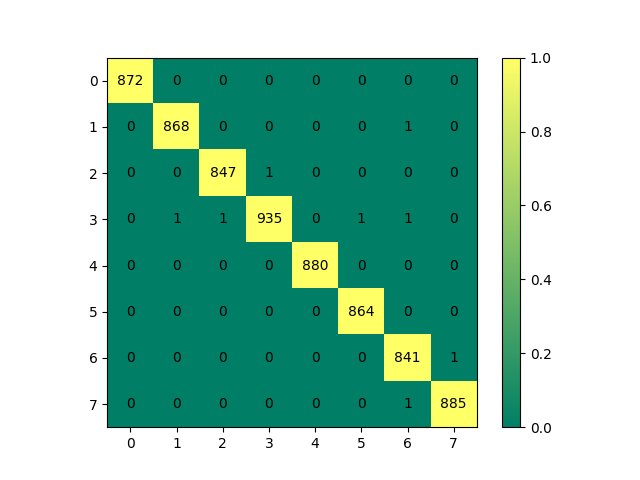}
\caption{Confusion matrix of EEG-S}
\label{fig:con_m_eegmmidb}
\end{minipage}
\end{figure}

\begin{figure}[ht]
\centering
\begin{minipage}[b]{0.32\linewidth}
\centering
\includegraphics[width=\textwidth]{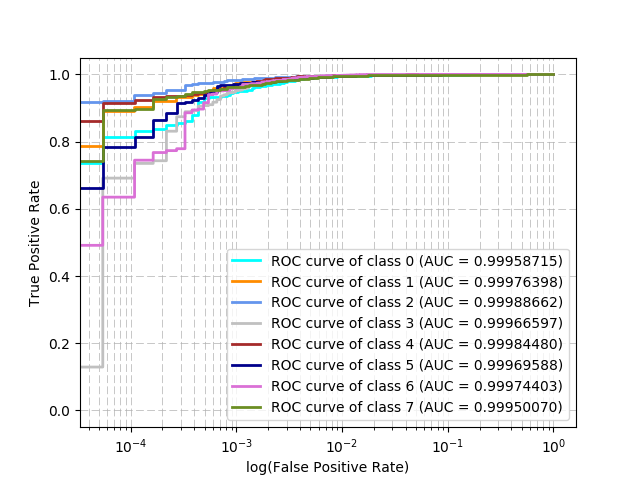}
\caption{ROC and AUC of EID-M}
\label{fig:ROC_3trials}
\end{minipage}
\begin{minipage}[b]{0.32\linewidth}
\centering
\includegraphics[width=\textwidth]{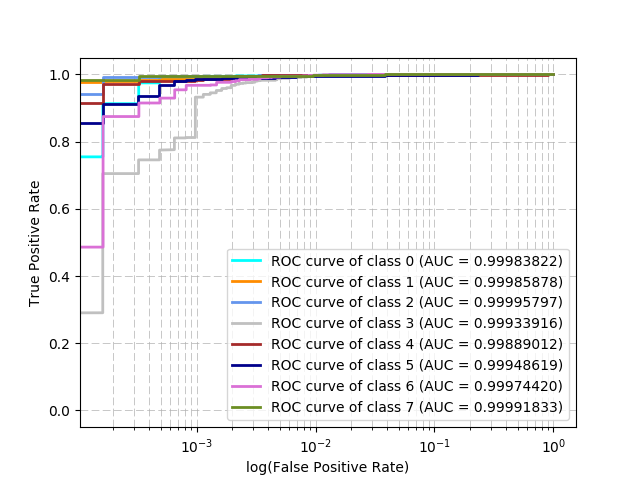}
\caption{ROC and AUC of EID-S}
\label{fig:ROC_1trial}
\end{minipage}
\begin{minipage}[b]{0.32\linewidth}
\centering
\includegraphics[width=\textwidth]{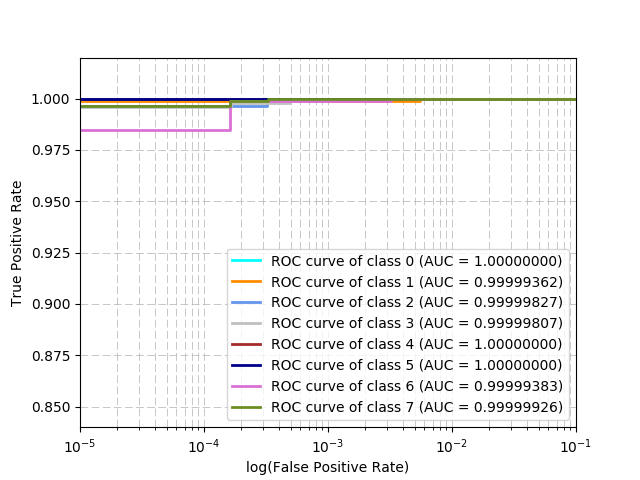}
\caption{ROC and AUC of EEG-S}
\label{fig:ROC_eegmmidb}
\end{minipage}
\end{figure}


\begin{table}[]
\centering
\caption{Evaluation report of EID-M dataset. The overall accuracy achieves 0.982 of 21000 testing samples. The support is the number of samples of each class.}
\label{tab:report_3trials}
\begin{tabular}{lllllllllc}
\rowcolor[HTML]{C0C0C0}
\hline
\textbf{} & \textbf{0} & \textbf{1} & \textbf{2} & \textbf{3} & \textbf{4} & \textbf{5} & \textbf{6} & \textbf{7} & \textbf{Average/Total} \\ \hline
\textbf{Precision} & 0.9723 & 0.9789 & 0.9777 & 0.9894 & 0.989 & 0.9814 & 0.9898 & 0.9774 & \textbf{0.982} \\
\textbf{Recall} & 0.9822 & 0.9885 & 0.9945 & 0.9711 & 0.9808 & 0.9821 & 0.9742 & 0.9834 & 0.9821 \\
\textbf{F1-score} & 0.9772 & 0.9837 & 0.9860 & 0.9802 & 0.9849 & 0.9818 & 0.9820 & 0.9804 & 0.982 \\
\textbf{Support} & 2674 & 2554 & 2601 & 2650 & 2639 & 2634 & 2636 & 2612 & 21000 \\ \hline
\end{tabular}
\end{table}


\begin{table}[]
\centering
\caption{The accuracy comparison with baselines and the state-of-the-art methods over EID-M dataset. The result shows that our approach achieves the highest accuracy of 0.982.}
\label{tab:comparison}
\begin{tabular}{llllll}
\rowcolor[HTML]{C0C0C0}
\hline
{\bf Index} & {\bf Method} & {\bf Acc} & {\bf Recall} & {\bf F1-Sore} & {\bf AUC} \\ \hline
1 & Jayarathne\cite{jayarathne2016brainid} & 0.919 & 0.914 & 0.9165 & 0.946 \\
2 & Bashar et al. \cite{bashar2016human} & 0.873 & 0.898 & 0.8853 & 0.907 \\
3 & Keshishzadeh et al. \cite{keshishzadeh2016improved} & 0.815 & 0.843 & 0.8288 & 0.859 \\
4 & Gui et al.\cite{gui2014exploring} & 0.833 & 0.811 & 0.8219 & 0.842 \\
5 & Thomas and Vinod \cite{thomas2016utilizing} & 0.859 & 0.869 & 0.8640 & 0.888 \\
6 & Kumari and Vaish \cite{kumari2015brainwave} & 0.875 & 0.872 & 0.8735 & 0.901 \\
7 & RF & 0.795 & 0.813 & 0.8039 & 0.827 \\
8 & KNN & 0.849 & 0.836 & 0.8424 & 0.847 \\
9 & RNN & 0.815 & 0.803 & 0.8090 & 0.821 \\
10 & RNN+XGB & 0.808 & 0.789 & 0.7984 & 0.803 \\
11 & PD+RNN & 0.853 & 0.821 & 0.8367 & 0.844 \\
12 & AR+RNN & 0.811 & 0.798 & 0.8044 & 0.831 \\
13 & XGB & 0.815 & 0.811 & 0.8130 & 0.853 \\
14 & PD+XGB & 0.965 & 0.959 & 0.9620 & 0.977 \\
15 & Ours (EID-M) & \textbf{0.982} & \textbf{0.9821} & \textbf{0.9820} & \textbf{0.999}\\ \hline
\end{tabular}
\end{table}

\subsection{Overall comparison} 
\label{sub:overall_comparison}

In this section, we firstly report our model's performance evaluated on EID-M dataset and then compare the proposed approach with the state-of-the-art approaches and baselines. Our approach
extract Delta wave through pattern decomposition fed it into an attention-based Encoder-Decoder RNN model, and predict the user's ID via a boosting classifier.
We randomly select 147,000 samples from EID-M to train the model and the residual 21,000 samples are used to test the performance. Through tuning, the hyper-parameters used in our approach are listed following. In EEG pattern decomposition, we employ a 3 order butter-worth band-pass filter and the passband is $[0.5Hz, 4Hz]$. In the attention-based RNN structure, the encoder consists of 1 input layer (14 nodes), 3 non-recurrent fully-connected hidden layers (164 nodes) and 1 recurrent LSTM layer (164 cells); the decoder includes 1 fully-connected hidden layer (164 nodes) and 1 output layer (8 nodes). The learning rate is 0.001; the parameter of $\ell-2$ norm is set as 0.001; the encoder and decoder separately have 6 and 2 layers; training dataset is divided into 7 batches with the batch size of 21,000; the number of training iterations is 2000. In the classifier: the learning rate is 0.7; the sub-sampling rate is 0.9; the max depth is set as 6; the training iterations is 500. The ground truth (from 0 to 7) is represented as a one-hot label which corresponding to the ID of subjects.

The proposed approach achieves the identification accuracy as {\bf 0.982}. The detailed confusion matrix, evaluation report, and ROC curves (with AUC scores) are illustrated in Figure~\ref{fig:con_m_3trials}, Table~\ref{tab:report_3trials}, and Figure~\ref{fig:ROC_3trials}, respectively. The above evaluation metrics illustrate that our approach obtains higher than 0.97 precision of each class.

In addition, the accuracy comparison between our method and other state-of-the-art and baselines are listed in Table~\ref{tab:comparison}. RF denotes Random Forest, AdaBoost denotes Adaptive Boosting, LDA denotes Linear Discriminant Analysis, PD denotes for Pattern Decomposition, AR denotes AutoRegressive method, and XGB denotes for X-Gradient Boosting classifier (the classifier used in our approach).
In addition, the key parameters of the baselines are listed here: Linear SVM ($C=1$), RF ($n=200$), KNN (k=3),
and AR (13 order autoregressive from 40 samples). The setting up of PD, RNN and XGB classifier are same as the hyper-parameters mentioned above.
The methods used in the state-of-the-art are introduced as follows:
\begin{itemize}
\item Jayarathne et al. \cite{jayarathne2016brainid} focus on the 8 to 30 Hz Alpha and Beta combined frequency band across all EEG channels and extract the Common Spatial Patterns (CSP) values as classification features. LDA is employed as the classifier.
\item Bashar et al. \cite{bashar2016human} first remove noise and artifacts using Bandpass FIR filter. Then learning the features through multi-scale shape description (MSD), multi-scale wavelet packet statistics (WPS) and multi-scale wavelet packet energy statistics (WPES). These features are finally used to train a support vector machine (SVM) classifier.
\item  Keshishzadeh et al. \cite{keshishzadeh2016improved} investigates the Autoregressive (AR) coefficients as the feature set which is identified by an SVM classifier.
\item Gui et al.\cite{gui2014exploring} propose to reduce the noise level through a low-pass filter, extract frequency features using wavelet packet decomposition, and perform classification based on a deep neural network.
\item Thomas and Vinod \cite{thomas2016utilizing} combines subject-specific alpha peak frequency, peak power, and delta band power values to form discriminative feature vectors and templates.
\item Kumari and Vaish \cite{kumari2015brainwave} apply discrete wavelet analysis to decompose EEG raw signal corresponding to EEG sub-band frequency (0-59Hz). The extracted statistical measures and energy calculation of each decomposed wave are classified by neural network structure.
\end{itemize}
All the approaches are working on the preprocessed EID-M dataset. The results show that our method achieves the highest accuracy of \textbf{0.982} compared with other methods.


\subsection{Efficiency Evaluation} 
\label{sub:efficiency_evaluation}
In this section, the efficiency refers to the required identification time. The low efficiency may limit the suitability for practical deployment. To assess the efficiency of the proposed approach, we focus on the algorithm running time and compared it with the widely used baselines and other classification methods. In this paper, we run the experiments on a GPU-accelerated machine with Nvidia Titan X pascal GPU, 768G memory, and 145 TB PCIe based SSD.

\begin{figure}
\centering
\begin{minipage}[b]{0.3\linewidth}
\centering
\includegraphics[width=\textwidth]{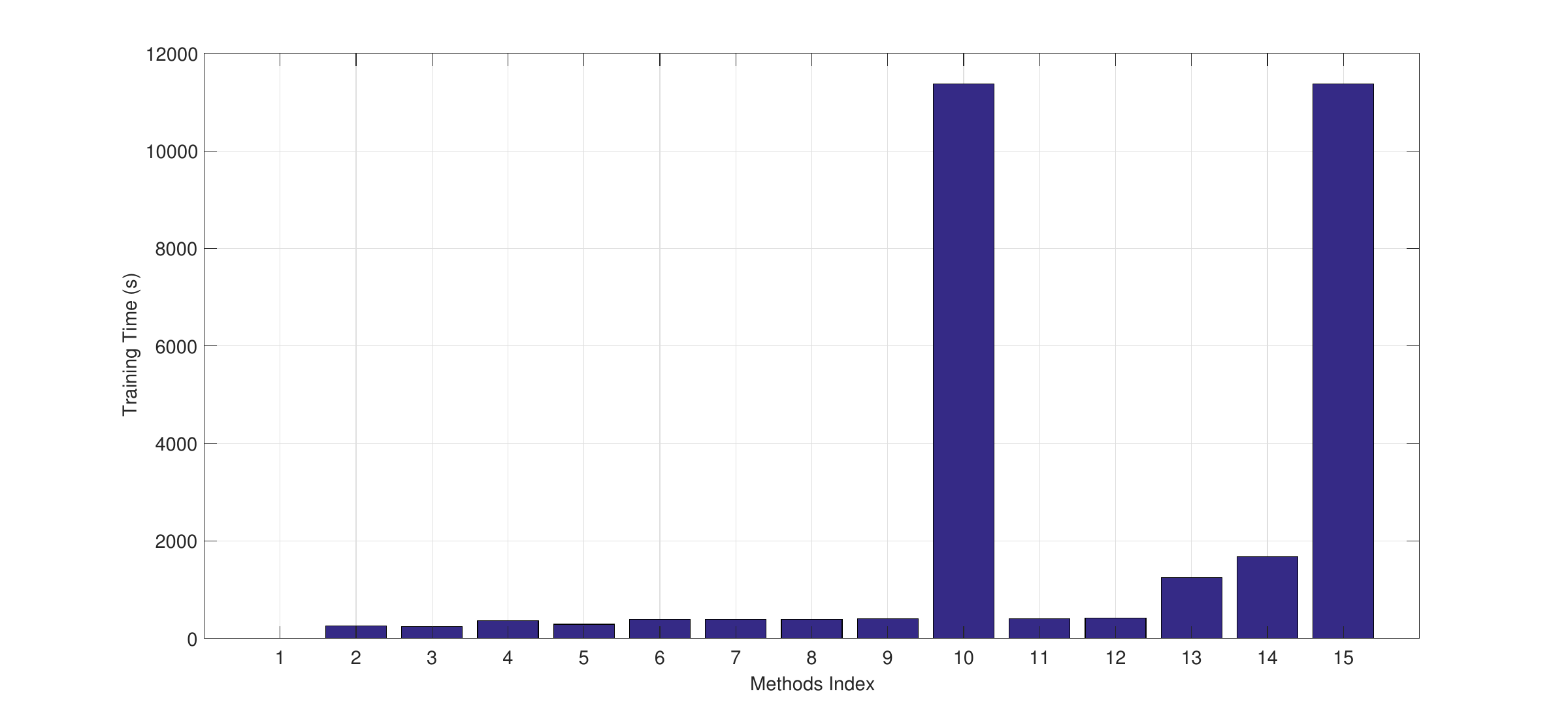}
\caption{Training time. The index corresponding the index in Table~\ref{tab:comparison}.}
\label{fig:training_time}
\end{minipage}
\begin{minipage}[b]{0.29\textwidth}
\includegraphics[width=\textwidth]{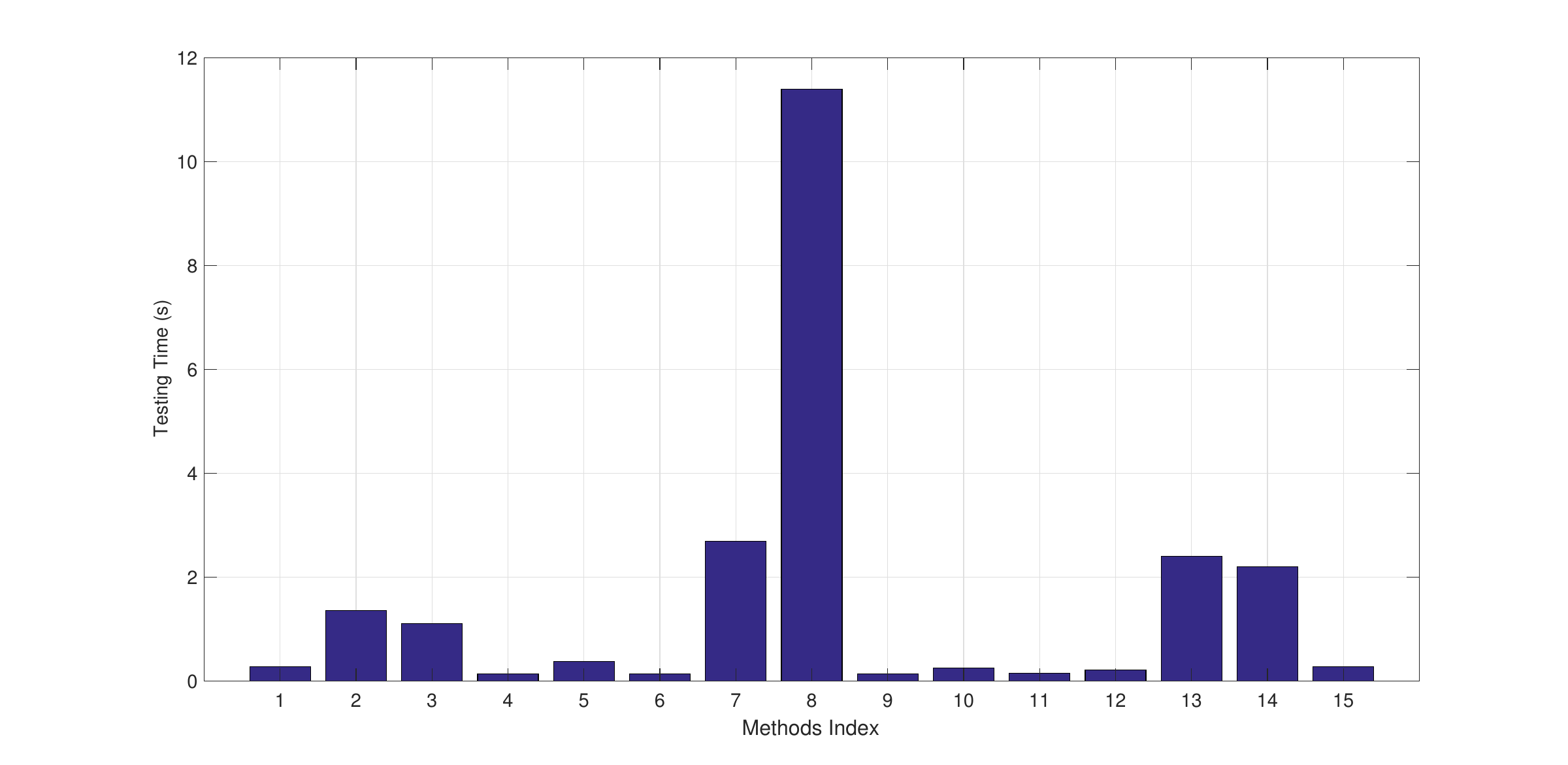}
\caption{Testing time. The index corresponding the index in Table~\ref{tab:comparison}.}
\label{fig:testing_time}
\end{minipage}
\begin{minipage}[b]{0.3\textwidth}
\includegraphics[width=\textwidth]{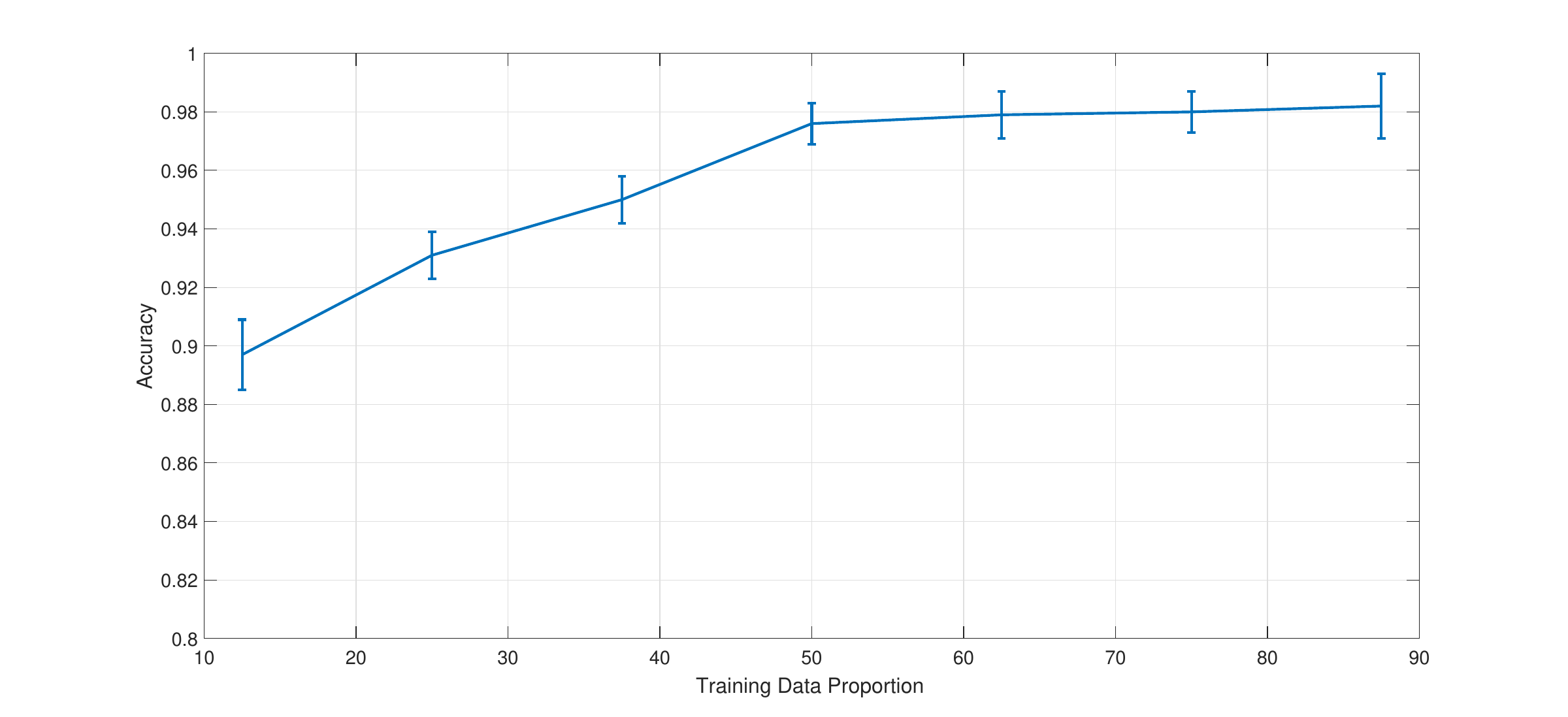}
\caption{The accuracy change trend with training data size }
\label{fig:datasize}
\end{minipage}%
\end{figure}

The time required to train the identification model is firstly given in Figure~\ref{fig:training_time}(the X-axis label denotes the index of algorithms shown in Table~\ref{tab:comparison}), which illustrates that our approach (PD+RNN+XGB) and the RNN+XGB approach take much more training time than other methods. The reasons are in two aspects. On one hand, the algorithm loops take a considerable amount of time (RNN run for 2000 iterations and XGB run for 500 loops). On the other hand, the deep learning structure and the boosting trees have much more parameters and complex structures than other classification models. Compared to the training time, however, for practical considerations, the execution time of an algorithm during testing is more important. Figure~\ref{fig:testing_time} presents that the testing time of our model is less than 1 second, which is shorter than most of the state-of-the-arts and baselines. Summarily, our model takes only tiny testing time although it requires more time to train the model, which is acceptable and reasonable in the real world implement.

In the practical deployment, the data size used to train the model generally is one important impact factor of the model's performance. We conduct a set of experiments to investigate the accuracy influence brought by training data size. We run the experiments for 5 times and report the error-bar of results in Figure~\ref{fig:datasize}, which shows that our approach could achieve the accuracy around 0.9 even when only 12.5\% of the available dataset is used for training. This presents that the proposed approach has a low dependency on the training data size.

\subsection{Robustness evaluation} 
\label{sub:robustness_evaluation}
In practical scenarios and real-world deployment, the identification system is applied to the multi-trial situation. The data used to train the system and the test data used to identify the user come from different trials (different placements of the device). For example, the user wears the EEG headset and collect the first trial data; then collect the second trial data after he/she removes the headset and puts it back again. There maybe some difference between two trials data, which is caused by the different placement position or other internal equipment reasons. Therefore, The divergence of the training data and testing data should be considered when the identification system is designed.

In this section, we evaluate the robustness of the proposed approach by analyzing how the single-trial/multi-trial affect the identification accuracy. Two datasets, which respectively contain single-trial identification data (EID-S) and multi-trial identification data (EID-M), are employed to assess our method. More details about the datasets are provided in Section~\ref{sub:setting}.

The evaluations of EID-S is shown in Table~\ref{tab:report_1trial}, through which we can observe that our approach achieves the overall accuracy of {\bf 0.9882\%} on EID-S and the precisions of all classes are higher than 0.98. To take a closer look at the result, confusion matrix (Table~\ref{fig:con_m_1trial}) and ROC curves (Figure~\ref{fig:ROC_1trial}) are provided. The performances of EID-M are reported in Section~\ref{sub:overall_comparison} (Figure~\ref{fig:con_m_3trials}, Table~\ref{tab:report_3trials}, and Figure~\ref{fig:ROC_3trials}). Through the comparison of the performances of EID-M and EID-S, we could know that the identification overall accuracy has a slight decrease (from 0.9882 to 0.982) with the increase of data trials. The inter-trial divergence only contributes a slight fluctuation (0.062) on the identification accuracy. This fact illustrates that the proposed approach has potential on the real world implement and large-scale application.

\begin{table}[]
\centering
\caption{Evaluation report of EID-S dataset. The overall accuracy achieves 0.9882 of 7000 testing samples.}
\label{tab:report_1trial}
\begin{tabular}{lllllllllc}
\rowcolor[HTML]{C0C0C0}
\hline
\textbf{} & \textbf{0} & \textbf{1} & \textbf{2} & \textbf{3} & \textbf{4} & \textbf{5} & \textbf{6} & \textbf{7} & \textbf{Average/Total} \\ \hline
\textbf{Precision} & 0.9897 & 0.9881 & 0.9944 & 0.9837 & 0.9895 & 0.9844 & 0.9866 & 0.9897 & \textbf{0.9882} \\
\textbf{Recall} & 0.992 & 0.9924 & 0.9944 & 0.9712 & 0.986 & 0.9939 & 0.9789 & 0.9977 & 0.9883 \\
\textbf{F1-score} & 0.9908 & 0.9903 & 0.9944 & 0.9774 & 0.9878 & 0.9891 & 0.9827 & 0.9937 & 0.9883 \\
\textbf{Support} & 872 & 927 & 892 & 857 & 857 & 831 & 893 & 871 & 7000 \\ \hline
\end{tabular}
\end{table}

\subsection{Adaptability evaluation} 
\label{sub:adaptability_evaluation}
To examine the adaptability and consistency, our model is evaluated on another dataset which is more precisely but difficult-to-operate. According to {\em the principle of single variable}, both the local dataset (EID-S) and the public dataset (EEG-S) are collected from single-trial and contains 56,000 samples belong to 8 subjects. The details of EID-S and EEG-S can be found in Section~\ref{sub:setting}. Compared with the Emotiv headset used in EID-S, the BCI 2000 system used in EEG-S is more accurately but inconvenient.

The experiment report (Table~\ref{tab:report_eegmmidb}) of EEG-S illustrates our model gains the accuracy of {\bf 0.9989} and all the evaluation metrics (precision, recall, and F1-score) are higher than 0.995. The confusion matrix and ROC curves are given in Figure~\ref{fig:con_m_eegmmidb} and Figure~\ref{fig:ROC_eegmmidb}, respectively. The accurate classification of EEG-S demonstrates that our approach has good adaptability and ables to handle different situations (like various EEG equipment).

Recall the results of EID-S (Figure~\ref{fig:con_m_1trial}, Table~\ref{tab:report_1trial}, and Figure~\ref{fig:ROC_1trial}), EEG-S performs better and achieves an accuracy of around 0.01 improvement. The reason is that EEG-S has more channels (64 vs 14)  and higher sampling rate ($160 Hz vs 128 Hz$) which encloses more useful information for the identification.

This section and the previous section illustrate that our approach has the potential to be largely deployment in practice environment from different aspects (robustness and adaptability).

\begin{table}[]
\centering
\caption{Evaluation report of EEG-S dataset. The overall accuracy achieves 0.9989 of 7000 testing samples.}
\label{tab:report_eegmmidb}
\begin{tabular}{lllllllllc}
\rowcolor[HTML]{C0C0C0}
\hline
\textbf{} & \textbf{0} & \textbf{1} & \textbf{2} & \textbf{3} & \textbf{4} & \textbf{5} & \textbf{6} & \textbf{7} & \textbf{Average/Total} \\ \hline
\textbf{Precision} & 1 & 0.9988 & 0.9988 & 0.9957 & 1 & 1 & 0.9988 & 0.9989 & \textbf{0.9989} \\
\textbf{Recall} & 1 & 0.9988 & 0.9988 & 0.9989 & 1 & 0.9988 & 0.9964 & 0.9989 & 0.9988 \\
\textbf{F1-score} & 1 & 0.9988 & 0.9988 & 0.9973 & 1 & 0.9994 & 0.9976 & 0.9989 & 0.9989 \\
\textbf{Support} & 872 & 869 & 848 & 939 & 880 & 864 & 842 & 886 & 7000 \\ \hline
\end{tabular}
\end{table}

\subsection{EEG pattern decomposition effects} 
\label{sub:eeg_decomposition_effects}
This section designs a set of comparison experiments to validate the hypothesis proposed in Section~\ref{sec:eeg_bands_analysis}, which claims that the Delta pattern signals takes the most distinguishable information for identification. To demonstrate the priority of Delta pattern, we decompose the EEG data into 6 patterns: Delta pattern, Theta pattern, Alpha pattern, Beta pattern, Gamma pattern, and Full-frequency pattern.
The Full-frequency pattern contains full frequency bands from 0 to 128 $Hz$. Note that the sampling rate in the local datasets is 128 $Hz$, which means that the maximum filtering range of butter-worth filter is 0-64 $Hz$. Therefore, the Gamma pattern used in this study is set as 30-63 $Hz$.

Our approach and other widely used classifiers are evaluated all of the EID-M, EID-S, and EEG-S datasets over 6 different patterns. The experiments results are shown in Table~\ref{tab:filter_comparison}. The primary conclusions are listed as follows:
\begin{itemize}
  \item Our approaches achieves the highest accuracy on all of the three datasets (with different trials, collection equipment, and sampling precision), which proofs that our model has outstanding robustness and adaptability.
  \item Delta pattern signals provide higher identification accuracy compared with other 5 categories of patterns over all datasets. This fact presents that Delta pattern contains the most discriminative information for identification and demonstrates the hypothesis proposed in Section~\ref{sec:eeg_bands_analysis} is appropriate.
  \item Several statistic based classification models (such as RF, KNN, and XGB) work well on the low-frequency patterns (Delta and Theta) but cannot handle high-frequency band signals (Alpha, Beta, and Gamma).
  \item The deep learning algorithm can extract deep relationships between samples from complicated and high fluctuate situations. This conclusion can be inferred from the observations that RNN has lower accuracy than RF/KNN/XGB in Delta and Theta patterns but performs better in other patterns.
  The above two attributes inspire the combination of the attention-based RNN structure and the tree-boosting classifier.
  \item The baselines and the state-of-the-art methods can gain acceptable identification accuracy on high-quality EEG dataset but fails on the low-quality dataset. Take the Full-frequency pattern as an example, RF/XGB/RNN achieves the accuracy of more than 0.95 on EEG-S but lower than 0.82 on EID-M. {\it However, our approach keeps consistently high accuracy no matter the data quality.} This phenomenon promotes the future deployment in practical of our approach.
\end{itemize}

\begin{table}[]
\centering
\caption{EEG Pattern Decomposition Analysis}
\label{tab:filter_comparison}
\begin{tabular}{llllllllc}
\rowcolor[HTML]{C0C0C0}
\hline
\textbf{Dataset} & \textbf{Methods} & \multicolumn{6}{c}{\cellcolor[HTML]{C0C0C0}\textbf{EEG Patterns}} & \textbf{Best Level} \\ \cline{3-8}
\rowcolor[HTML]{C0C0C0}
\textbf{} & \textbf{} & \textbf{Delta} & \textbf{Theta} & \textbf{Alpha} & \textbf{Beta} & \textbf{Gamma} & \textbf{Full} & \textbf{} \\ \hline
 & \textbf{SVM} & 0.143 & 0.157 & 0.137 & 0.135 & 0.138 & 0.2745 &  \\
 & \textbf{RF} & 0.936 & 0.707 & 0.677 & 0.489 & \textbf{0.435} & 0.7935 &  \\
 & \textbf{KNN} & 0.941 & \textbf{0.804} & 0.618 & 0.35 & 0.313 & 0.819 &  \\
 & \textbf{AdaBoost} & 0.251 & 0.13 & 0.15 & 0.15 & 0.171 & 0.24 &  \\
 & \textbf{LDA} & 0.148 & 0.154 & 0.135 & 0.135 & 0.129 & 0.28 &  \\
 & \textbf{XGB} & 0.965 & 0.665 & 0.69 & 0.495 & 0.414 & 0.815 &  \\
 & \textbf{RNN} & 0.917 & 0.709 & 0.708 & 0.518 & 0.411 & 0.813 &  \\
\multirow{-8}{*}{\textbf{EID-M}} & \textbf{Ours} & \cellcolor[HTML]{FCFF2F}\textbf{0.982} & 0.713 & \textbf{0.73} & \textbf{0.513} & 0.423 & \textbf{0.822} & \multirow{-8}{*}{\textbf{\begin{tabular}[c]{@{}c@{}}0.982 \\ (Delta)\end{tabular}}} \\ \hline
 & \textbf{SVM} & 0.135 & 0.162 & 0.181 & 0.152 & 0.132 & 0.408 &  \\
 & \textbf{RF} & 0.947 & 0.771 & 0.719 & 0.587 & 0.377 & 0.863 &  \\
 & \textbf{KNN} & 0.953 & \textbf{0.824} & 0.714 & 0.472 & 0.495 & 0.853 &  \\
 & \textbf{AdaBoost} & 0.278 & 0.29 & 0.162 & 0.2 & 0.16 & 0.3 &  \\
 & \textbf{LDA} & 0.14 & 0.16 & 0.183 & 0.152 & 0.122 & 0.41 &  \\
 & \textbf{XGB} & 0.981 & 0.785 & 0.791 & 0.599 & 0.489 & 0.893 &  \\
 & \textbf{RNN} & 0.9425 & 0.7568 & 0.8175 & \textbf{0.6331} & 0.5141 & 0.9045 &  \\
\multirow{-8}{*}{\textbf{EID-S}} & \textbf{Ours} & \cellcolor[HTML]{FCFF2F}\textbf{0.9882} & 0.821 & \textbf{0.8259} & 0.612 & \textbf{0.517} & \textbf{0.913} & \multirow{-8}{*}{\textbf{\begin{tabular}[c]{@{}c@{}}0.9882 \\ (Delta)\end{tabular}}} \\ \hline
 & \textbf{SVM} & 0.216 & 0.167 & 0.148 & 0.169 & 0.186 & 0.652 & \\
 & \textbf{RF} & 0.972 & 0.885 & 0.819 & 0.823 & 0.87 & 0.957 &  \\
 & \textbf{KNN} & 0.974 & 0.865 & 0.781 & 0.559 & 0.743 & 0.936 &  \\
 & \textbf{AdaBoost} & 0.32 & 0.32 & 0.27 & 0.23 & 0.22 & 0.34 &  \\
 & \textbf{LDA} & 0.186 & 0.17 & 0.28 & 0.168 & 0.162 & 0.6618 &  \\
 & \textbf{XGB} & 0.9972 & \textbf{0.982} & 0.967 & 0.959 & 0.953 & 0.989 &  \\
 & \textbf{RNN} & 0.9981 & 0.9667 & 0.964 & 0.947 & 0.952 & 0.9886 &  \\
\multirow{-8}{*}{\textbf{EEG-S}} & \textbf{Ours} & \cellcolor[HTML]{FCFF2F}\textbf{0.9989} & 0.972 & \textbf{0.968} & \textbf{0.961} & \textbf{0.955} & \textbf{0.99} & \multirow{-8}{*}{\textbf{\begin{tabular}[c]{@{}c@{}}0.9989 \\ (Delta)\end{tabular}}} \\ \hline
\end{tabular}
\end{table}


\section{Discussion and future work} 
\label{sec:discussion}
In this paper, we propose an EEG-based identification approach and evaluate the robustness and adaptability over three datasets. In this section, we discuss the challenges and potential future work of our research.

First of all, the impaction of multi-trial worth to attract more attention although we have investigated a preliminary study on this topic. Limited by the local experimental conditions, our study only gathered EEG data from 8 subjects with few trials. The dataset is only divided into two categories (Multi and Single), which is not enough to explore the change trend of the identification accuracy with the increase of data trials. The accuracy trend is supposed to be investigated over the dataset with enough trials.

Moreover, the pre-trained model should be updated for a period of time since the user's EEG data is gradually changed with the environmental factors such as age, mental state, and living style. One of our future work is to develop an online learning system which is enabled to automatically update the training dataset based on the testing data which is collected during the operating period.

In addition, the emotional threshold is one potential challenge faced by user identification. It is well known that the EEG signals are associated with user's emotion. Therefore, an emotional threshold is required to tolerate the slight emotion fluctuation which may be caused by routine factors such as fatigue and temporal emotion shift. At the same time, the threshold is demanded to detect the out-of-bound emotions which may be occurred in dangerous situations such as being hacked by a terrorist.

\section{Conclusion} 
\label{sec:conclusion}
Taking the advantages of EEG-based techniques for attack-resilient, we propose a biometric EEG-based identification approach, to overcome the limitations of traditional biometric identification methods. We analyzed the EEG data pattern characteristics and capture the Delta pattern which takes the most distinguishable features for user identification. Based on the pattern decomposition analysis, we report the structure of the proposed approach. In the first step of identification, the preprocessed EEG data is decomposed into Delta pattern. Then an attention-based RNN structure is employed to extract deep representations of Delta wave. At last, the deep representations are used to directly identify the user' ID. The proposed approach is evaluated over 3 datasets (two local and one public dataset). The experiments results illustrate that our model achieves the accuracy of $0.982$, $0.9882$, and $0.9989$ over three datasets, separately. The results also infer the robustness and adaptability of our model. Moreover, a set of experiments are conducted and verified that the Delta pattern is the most reliable and dominant pattern in EEG-based identification.


\begin{acks}

\end{acks}

\bibliographystyle{ACM-Reference-Format}
\bibliography{sigproc.bib}

\end{document}